# Photon random walk model of low-coherence enhanced backscattering (LEBS) from anisotropic disordered media: a Monte Carlo simulation

Hariharan Subramanian, Prabhakar Pradhan, Young L. Kim, Yang Liu, Xu Li and Vadim Backman

*Department of Biomedical Engineering, Northwestern University, Evanston, IL 60208.*

Constructive interference among coherent waves traveling time-reversed paths in a random medium gives rise to the enhancement of light scattering observed in directions close to backscattering. This phenomenon is known as enhanced backscattering (EBS). According to diffusion theory, the angular width of an EBS cone is proportional to the ratio of the wavelength of light $\lambda$ to the transport mean free path length $l_s^*$ of a random medium. In biological media, large $l_s^* \sim$ 0.5-2 mm $\gg \lambda$ results in an extremely small (~0.001°) angular width of the EBS cone making the experimental observation of such narrow peaks difficult. Recently, the feasibility of observing EBS under low spatial coherence illumination (spatial coherence length $L_{sc} \ll l_s^*$) was demonstrated. Low spatial coherence behaves as a spatial filter rejecting longer path-lengths and, thus, resulting in more than 100 times increase in the angular width of low coherence EBS (LEBS) cones. However, conventional diffusion approximation-based model of EBS has not been able to explain such dramatic increase in LEBS width. Here we present a photon random walk model of LEBS using Monte Carlo simulation to elucidate the mechanism accounting for the unprecedented broadening of LEBS peaks. Typically, the exit angles of the scattered photons are not considered in modeling EBS in diffusion regime. We show that small exit angles are highly sensitive to low order scattering, which is crucial for accurate modeling of LEBS. Our results show that the predictions of the model are in excellent agreement with experimental data.

## 1. Introduction

The constructive self-interference effect due to the coherent waves traveling in time-reversed paths in a disordered medium produces an enhanced intensity cone in directions close to the backscattering. In case of complete diffusion of light, the amplitude of this intensity profile as a function of the backscattering angle can be as high as twice that of the incoherent background.[1] This increase in reflectivity in the backward direction leads to a reduction in the amount of light transported in the forward direction thereby resulting in the weak localization of photon. This phenomenon of enhanced backscattering (EBS, also known as coherent backscattering) was first theoretically shown by Kuga and Ishimaru,[1] and experimentally observed by Lagendijk and colleagues.[2] Thereafter, EBS has become a subject of major research interest.[3,4] In order to explain the enhanced backscattering, typically the light is assumed to be completely spatially coherent.[5,6] In a homogeneous semi-infinite disordered medium, the full angular width at half-maximum (FWHM), $\omega_{hm}$, of the EBS cone was shown to be inversely proportional to the ratio of the wavelength of light $\lambda$ to the transport mean free path length $l_s^*$ of light in the medium:[2,3]

$$\omega_{hm} = \lambda / (3\pi l_s^*). \qquad (1)$$

Although EBS enhancement has been widely studied in a variety of disordered media with relatively short $l_s^*$,[7-9] investigation of EBS in weakly scattering media with $l_s^* \gg \lambda$ has been exceedingly difficult, in part due to very small widths of EBS peaks predicted in such media (e.g., $\omega_{hm} \sim$ 0.001° for $l_s^* \sim$1 mm) and excessive speckle.[5,10,11] Only recently, in pioneering experiments, Yodh and Sapienza[12] have achieved detection of such narrow EBS peaks.

In particular, a biological tissue is one important example of a weakly scattering medium with long $l_s^*$. Measurement of light scattering and absorption properties of tissue is crucial to exploit the use of light for both diagnostic and therapeutic purposes.[13-20] Accordingly, EBS may be used as one of the potential tools for noninvasive optical characterization of tissue. However, only very few studies[10,21,22] have actually attempted EBS measurements in tissue. In particular, Alfano *et al.*[21,22]

first reported EBS in biological tissue using femtosecond-time-resolved measurements.

Recently, we demonstrated the feasibility of observing EBS under low spatial coherence illumination (spatial coherence length $L_{sc}<<l_s^*$). Low spatial coherence behaves as a spatial filter that rejects longer path-lengths with exponentially low probability, thus, resulting in more than 100 times increase in the angular width of low coherence EBS (LEBS) cones.[23,24] Furthermore, we showed that not only does LEBS represent a novel enhanced backscattering phenomenon, but it also opens up the feasibility of studying enhanced backscattering in biological tissue and other media with long $l_s^*$ and enables depth-selective spectroscopic tissue characterization.[25] For example, we demonstrated that LEBS can be used to diagnose the earliest, previously undetectable stage of colon carcinogenesis that precedes currently histologically detectable lesions.[23,24] These results underline the need for thorough understanding of this new effect.

Previously, LEBS was observed by combining EBS measurements with low spatial coherence (LSC) illumination and low temporal coherence detection. This technique uses a broadband continuous wave xenon lamp to achieve low spatial coherence illumination. We also demonstrated that the angular width of an EBS peak observed under LSC illumination (~0.3°) is more than 100 times broader than that of conventional EBS.[23] We note that one of the most intriguing properties of LEBS is the dramatically increased angular width of the LEBS peaks, which cannot be explained on the basis of conventional diffusion approximation based model of EBS alone. In order to further our understanding of this unprecedented broadening of LEBS peaks and identify the origin of LEBS, it is necessary to develop a rigorous model of LEBS.

In this paper we present the photon random walk model of low coherence enhanced backscattering using Monte Carlo simulations, which is subsequently compared with the experimentally obtained LEBS peaks. Monte Carlo simulations have been extensively used to simulate light propagation in biological tissue.[26,27] Many groups[28-31] have used Monte Carlo modeling of EBS peaks in biological and non-biological samples. Kaiser and colleagues[28] reported the first quantitative comparison between the experimentally observed EBS peaks and Monte Carlo simulation in cold atoms by taking into account the shape of atomic cloud and its internal structure. Delpy and colleagues[31] used Monte Carlo model of backscattered light from turbid media to simulate weak localization in biological tissues and were able to extract optical parameters such as scattering and absorption coefficients from angular intensity profiles of EBS peaks. Berrocal et al.[32] recently characterized intermediate scattering in sprays and other industrially relevant turbid media using Monte Carlo simulation. Specifically, they explained the influence of exit angle of photons on the relative intensity of different orders of scattering in the intermediate, single-to-multiple scattering regime and validated their results by Monte Carlo simulation.

In this paper, we model for the first time LEBS using Monte Carlo simulations, show that the model is in excellent agreement with experimental data, and explain the origin of LEBS broadening. We demonstrate that the exit angle of the scattered photons is of critical importance when the spatial coherence length of the light source is much smaller than $l_s^*$. On the other hand, we show that the exit angle of photons plays only a minimal role in the simulation of conventional EBS peaks.

The paper is organized as follows: Section 2 describes the theory of LEBS peaks and Monte Carlo simulations used in modeling the LEBS profile from a random dielectric medium. Section 3 addresses the importance of exit angle of photons in low order scattering regime compared with the diffusive regime. Section 4 explains the validation of our Monte Carlo model with analytic expression and the comparison with experimental results. Finally, in Section 5, we discuss and conclude our results.

## 2. Monte Carlo model of LEBS

In order to model low-coherence enhanced backscattering, we developed a photon random walk model using Monte Carlo (MC) simulation. EBS originates from constructive interference between any given light path and its time reversed counterpart related by the reciprocity theorem. MC simulation provides a distribution of the photon backscattering intensity as a function of radial distance and exit angle for relevant optical parameters. The shape of the EBS cone versus the scattering angle is calculated from the Fourier transform of the radial intensity distribution on the surface of the sample illuminated by a point source simulated by the Monte Carlo method. This technique has been successfully used in modeling EBS in non-biological and biological samples.[30,31] The intensity of the EBS cone thus obtained in the backward direction, $I_{EBS}$ can be written as,

$$I_{EBS}(\vec{q}_\perp) = \iint P(r)\exp(iq_\perp . r)d^2 r , \quad (2)$$

where $P(r)$ is the probability of radial intensity distribution of EBS photons with the radial vector $r$ perpendicular to the incident light and $q_\perp$ is the projection of wave vector onto the orthogonal plane in the backward direction. In an isotropic disordered medium, the two dimensional Fourier integral can be further simplified to,[24]

$$I_{EBS}(\vec{q}_\perp) \propto \int rP(r)\exp(iq_\perp . r)dr , \quad (3)$$

where $rP(r)$ is the radial intensity distribution of the conjugated time-reversed paths around a point-like light source illuminating a sample and the projection of wave vector, $q_\perp = 2\pi\theta/\lambda$.

In contrast to conventional EBS, low coherence enhanced backscattering is observed using a broadband light source with low spatial coherence.[23] Therefore, in order to model the effect of low spatial coherence length of illumination on EBS, we incorporated additional coherence length dependent weighting factor to the numerical simulation. We used the readily derived form of the degree of spatial coherence $C_{L_{sc}}(r)$ as follows:[34]

$$C_{L_{sc}}(r) = 2J_1(r/L_{sc})/(r/L_{sc}) , \quad (4)$$

where $J_1$ is the first order Bessel function, $L_{sc}$ is the spatial coherence length corresponding to the 88th percentile of the ideal value of unity, and $r$ is the radial vector perpendicular to the incident light. Thus, the modified LEBS intensity in the presence of a low coherence source can be written as

$$I_{LEBS}(\theta) \propto \int rP(r).C_{L_{sc}}(r)\exp(i\frac{2\pi\theta}{\lambda}r)dr . \quad (5)$$

The first term in the above equation, $rP(r) (\equiv I(r))$, is the radial intensity distribution, of the point like source illuminating the sample, which can be obtained using the Monte Carlo simulation.

A detailed description of the Monte Carlo simulation is given in references.[35,36] Here we only briefly describe the essential aspects of this method. In our simulations, approximately $10^{10}$ photon packets are launched into the sample. The optical parameters of the sample are assigned based on the sample size used in the LEBS experiment.[23] The samples are slabs with infinite lateral extents and having optical parameters of biological relevance ($l_s^*$ 500 μm -- 2000 μm, g ~ 0.6 -- 0.9, λ = 520 nm). The direction of incidence is normal to the sample's surface (*xy*-plane) with an initial weight $w$. As the specular reflectance is completely avoided in the experiment, the photon packets are allowed to propagate within the sample with its initial weight of $w$ and random step size $s$, given as $s = -\ln(\xi)/\mu_t$, where $\mu_t$ is the total interaction coefficient of the medium and ξ is the pseudorandom number uniformly distributed between 0 and 1. Once the photon packet reach the interaction site, a scattering direction is then defined by the deflection angle $\theta$ $(0<\theta<\pi)$ and azimuthal angle $\phi$ $(0<\phi<2\pi)$ that are statistically sampled. The probability distribution of the cosine of the deflection angle is chosen according to the Henyey-Greenstein anisotropic phase scattering probability function[37] given as

$$P_{HG}(\cos\theta) = (1-g^2)/(2(1+g^2-2g\cos\theta)^{3/2}) , \quad (6)$$

where $<\cos(\theta)> = g$. We assume the angular distribution to be azimuthally symmetrical, i.e. uniform distribution for $\phi$ $( = 2\pi\xi)$, such that, $P_\phi(\phi) = 1/(2\pi)$. The photon packet is terminated using Russian roulette technique[38] after it undergoes series of scattering events. The weight of the photon packets escaping from the medium in the forward or backward direction, is then recorded in a user defined grid system.

Our study is focused on the intensity profile of photons in the backward direction from low order scatterings. The radial and angular resolution of the grid system used in the Monte Carlo simulation to collect the low order scattering photons is specified as 1μm and 0.3˚ respectively. The optical parameters, such as the scattering coefficients and transport mean free path of the scattering medium, were obtained from the sample used in the experiment.

Once the reflectance probability $P(r)$ is calculated from the Monte Carlo simulation, Eq. (5) is then used to obtain the LEBS peak using the Fourier transform of the radial intensity distribution. The LEBS peak obtained is then convoluted with the angular response of the instrument (~ 0.04° – 0.3°) to compensate for the finite point-spread function of the detection system and the slight divergence of the incident beam. The width of the convoluted LEBS peak $W$ is defined as follows,[39]

$$W = \left(\int_0^\infty I_{LEBS}(\vec{q}_\perp)d\vec{q}_\perp\right)^2 / \left(\int I^2_{LEBS}(\vec{q}_\perp)d\vec{q}_\perp\right). \quad (7)$$

According to our simulations, $W$ provides a better metric to characterize LEBS peak width compared to FWHM. The width obtained from the simulation is then compared with those obtained from the LEBS experiment.

## 3. Results

### A. Effect of exit angle of photons in low orders of scattering

We first obtain the radial intensity distributed on the surface of the sample with $l_s^* = 2$ mm. The absorption coefficient $\mu_a$ and scattering coefficient $\mu_s$ are fixed at 0.0001 mm$^{-1}$ and 5 mm$^{-1}$ (at $\lambda = 520$ nm) respectively. The radial intensity distribution is calculated from the Monte Carlo simulation at different exit angles between 0.3° and 90°. The intensity profile at 3 different exit angles (1.5°, 20° and 80°) is shown in Fig. 1. Typically, in diffusion regime, the EBS peak is obtained directly from the Fourier transform of the radial intensity without taking into account the exit angle of the photons. From Fig. 1, it can be seen that the intensity profile for diffusive multiple scattering remains constant for different exit angles. This intensity profile obtained above can be written as,

$$\int_{r=0}^\infty \int_{\theta=0}^\pi p(r,\theta)drd\theta = N, \quad (8)$$

(for $\mu_a = 0$ and transmission = 0), where $N$ is the total number of photons injected. Let us now redefine the above probability integral for any finite angle as follows:

$$\int_{r=0}^\infty \int_{\theta=0}^{\theta_{ci}} p(r,\theta)drd\theta = \int_{r=0}^\infty p(r,\theta_{ci})dr = N_{\infty,ci}. \quad (9)$$

Now we normalize the above probability along the radial direction keeping the maximum angles $\theta_{ci}$ (i=1,2, 3,..) as fixed.

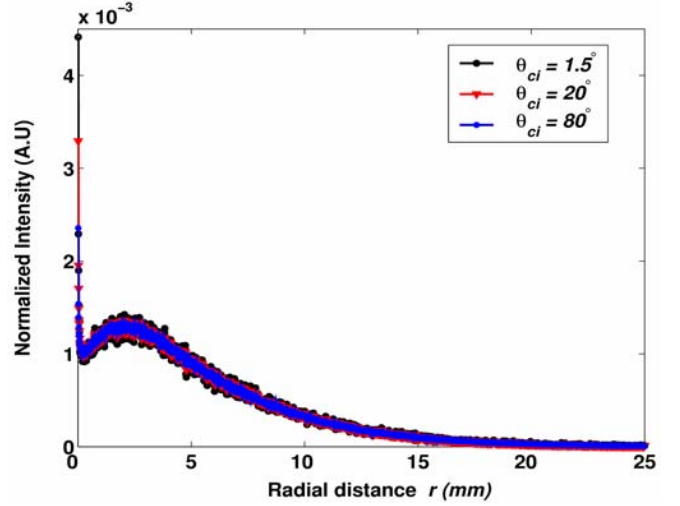

FIG. 1. Normalized intensity at different exit angles, $\theta_{ci}$ in the diffusive multiple scattering regime ($L_{sc} >> l_s^*$) is plotted as a function of radial distance $r$. The intensities are calculated using Monte Carlo simulation from a medium with $l_s^* = 2$ mm, $g = 0.9$ (at $\lambda = 520$ nm) and $L_{sc} = 50$ mm. The intensity profiles for different exit angles remains constant in the diffusive multiple scattering regime.

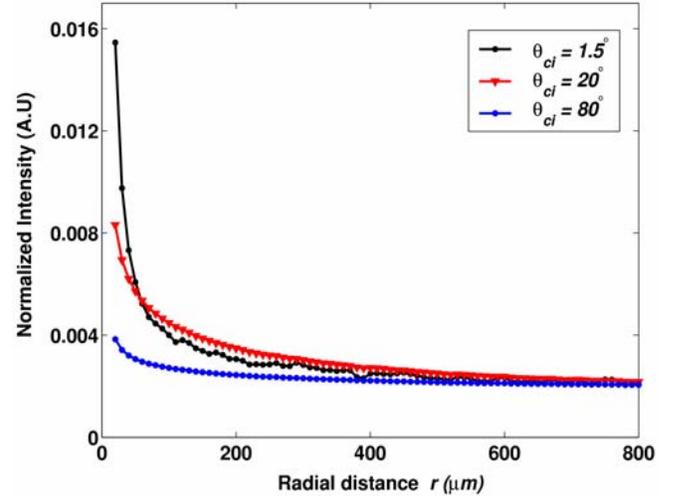

FIG. 2. Normalized intensity at different exit angles, $\theta_{ci}$ in the low order scattering regime ($L_{sc} << l_s^*$) is plotted as a function of radial distance $r$. The intensities are calculated using Monte Carlo simulation ($l_s^* = 2$ mm, $g = 0.9$, at $\lambda = 520$ nm) with $L_{sc} = 600$ μm. When the number of scattering events is restricted due to the finite spatial coherence area using low spatial coherence illumination, the intensity profile over $r$ becomes broader as $\theta_{ci}$ increases from 1.5° to 80°.

Let, $p(r,\theta_{ci})/N_{\infty,ci} = P(r,\theta_{ci})$, thus,

$$\int_0^\infty P(r,\theta_{ci})dr = 1. \qquad (10)$$

We now consider three different angles, $\theta_{c1} = 1.5°$, $\theta_{c2} = 20°$, and $\theta_{c3} = 80°$. The shape of the $P(r,\theta_{ci})$, (i =1,2, and 3) curves remains constant, which is shown in Fig. 1. However, a different picture emerges in the case of low spatial coherence illumination when the number of scattering events is restricted using low spatial coherence light source. $P(r,\theta_{ci})$ becomes broader as $\theta_{ci}$ increases from $1.5°$ to $80°$, as shown in Fig. 2.

In case of the spatial coherence light source the above probability distribution can be written as,

$$\int_{r=0}^{Lsc}\int_{\theta=0}^{\pi} p(r,\theta)drd\theta = N_{Lsc}, \qquad (11)$$

where $N_{Lsc}$ is the number of photons restricted by finite spatial area using low spatial coherence illumination, and,

$$\int_0^{Lsc}\int_0^{\theta ci} p(r,\theta)drd\theta = \int_0^{Lsc} p_{Lsc}(r,\theta_{ci})dr = N_{Lsc,ci}. \qquad (12)$$

Let, $p_{Lsc}(r,\theta_{ci})/N_{Lsc,ci} = P_{Lsc}(r,\theta_{ci})$, thus,

$$\int_0^{Lsc} P_{Lsc}(r,\theta_{ci})dr = 1. \qquad (13)$$

We point out that contrary to the diffusion regime, the shapes of $P(r,\theta_{ci})$ are now different. That is,

$\int_0^{Lsc} P_{Lsc}(r,\theta_{c1})dr$ is much narrower than $\int_0^{Lsc} P_{Lsc}(r,\theta_{c3})dr$.

For small radial distances corresponding to only few scattering events, it can be shown that,

$$\int_0^{Lsc/10} P_{Lsc}(r,\theta_{c1})dr >> \int_0^{Lsc/10} P_{Lsc}(r,\theta_{c3})dr. \qquad (14)$$

This result indicates the importance of considering the exit angles of photons particularly for modeling low coherence enhanced backscattering peaks.

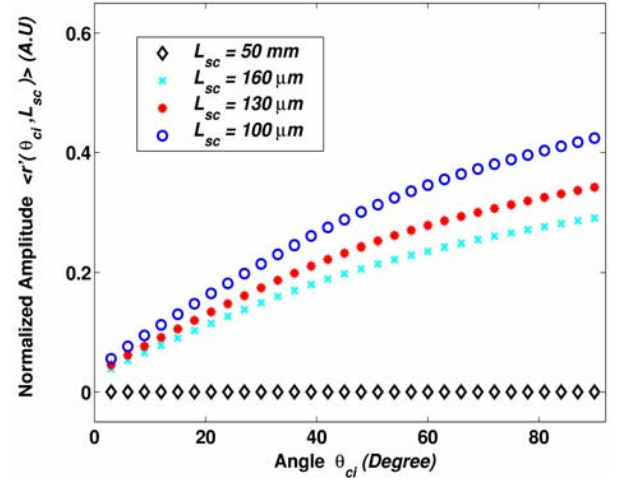

FIG. 3. $<r'(\theta_{ci},L_{sc})>$ as a function of exit angle $\theta_{ci}$ for four different $L_{sc}$. $<r'(\theta_{ci},L_{sc})>$ is simulated for a sample with $l_s^* = 2$ mm, and $g = 0.9$ (at $\lambda = 520$ nm) for different exit angles $\theta_{ci}$ varying from $1°$ to $90°$. $<r'(\theta_{ci},L_{sc})>$ is insensitive to the exit angle $\theta_{ci}$ when $L_{sc} >> l_s^*$, while $<r'(\theta_{ci},L_{sc})>$ increases with the increase in the exit angle in the low order scattering regime ($L_{sc} << l_s^*$).

As discussed above, the exit angles are less sensitive in the multiple scattering regime providing similar EBS peaks for different exit angles. In diffusion regime, the slight changes in reflectance probability $P(r)$ obtained from Monte Carlo simulation at different exit angles are translated over a radial distance of around 50 mm. Hence these small changes do not translate into narrow/broad peak when the Fourier transform of $P(r)$ is performed. In order to explain this fact we consider a function,

$$<r'(\theta_{ci},L_{sc})> = \int \frac{rP_{Lsc}(r,\theta_{ci})dr}{L_{sc}}. \qquad (15)$$

As expected, Fig. 3 shows that, $<r'(\theta_{ci},L_{sc})>$ remains constant with the change in exit angle in the EBS regime ($L_{sc} = 50$ mm $>> l_s^*$). On the other hand, in low coherence regime ($L_{sc} = 200$ μm $<< l_s^*$), we expect to see a change in the above function $<r'(\theta_{ci},L_{sc})>$ as the changes in $P(r)$ for different angles of collection are only translated over a very small radial distance. This effect is illustrated in Fig.3, which shows that in low coherence regime, $<r'(\theta_{ci}, L_{sc})>$ increases with the increase in exit angle. It is also shown that the slope of the curve increases as the spatial coherence length $L_{sc}$ decreases from 160 μm to 100 μm. Hence in low coherence regime, obtaining $P(r)$ for different angles of collection significantly affects the width of LEBS peaks. Therefore, it is imperative to obtain a proper exit angle to accurately model the LEBS peak in the low order scattering regime. This aspect is further explored in the subsequent subsections.

## B. Importance of proper exit angle in accurate modeling of LEBS peaks

The previous section clarifies that the exit angle of photons in the image plane is of critical importance for low coherence enhanced backscattering as LEBS probes low orders of scattering in a diffusive multiple scattering medium. This is due to the fact that the trajectories of the photons are clustered into a compact locus to exit at small angles for low orders of scattering, while the trajectories of the photons are less compact for higher orders of scattering. As can be seen from our simulations, the exit angle of the photons is sensitive to the depth from which the LEBS measurements are obtained. This fact is further illustrated in Figs. 4a and 4b, which show $<r(\theta_{ci}, L_{sc})>$ calculated for different exit angles and different spatial coherence length $L_{sc}$.

$$<r(\theta_{ci}, L_{sc})> = \int_0^{L_{scj}} r P_{Lsc}(r, \theta_{ci}) dr, \quad (16)$$

where $i = 0$ to 90 degrees and $j = 20$ to 140 μm.

As shown in Fig. 4a, for a given $L_{sc}$, $<r(\theta_{ci}, L_{sc})>$ increases with the increase of the exit angle from 0 to 90 degrees. Similarly, for a fixed exit angle, increase in $L_{sc}$ from 20 to 140 μm leads to an increase in $<r(\theta_{ci}, L_{sc})>$ as also shown in Fig. 4b. The relationship between $<r(\theta_{ci}, L_{sc})>$ and the depth of penetration of the scattered photons are obtained using Monte Carlo simulation by tracking the propagation of photons along the $z$-direction. The simulation is performed in the low order scattering regime and the size of the grid tracking the photons in the $z$-direction is kept at 1 μm. Figure 5 shows this relationship which indicates that $<r(\theta_{ci}, L_{sc})>$ is proportional to the depth from which the photons are scattered in the backward direction.

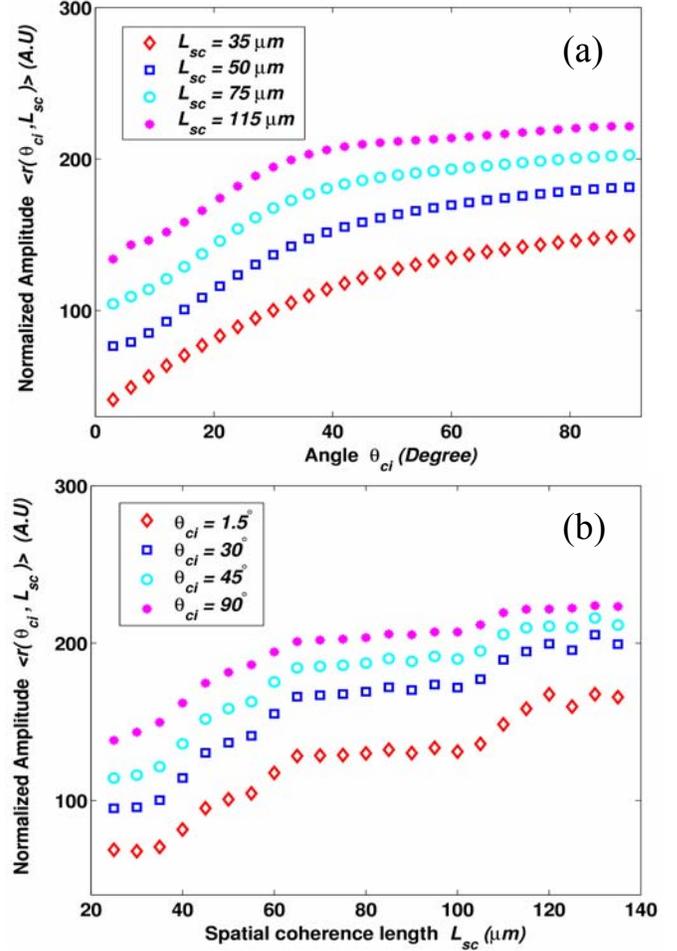

FIG. 4. (a) $<r(\theta_{ci}, L_{sc})>$ as a function of angle $\theta_{ci}$ obtained for 4 different $L_{sc}$. (b) $<r(\theta_{ci}, L_{sc})>$ as a function of $L_{sc}$ obtained for 4 different $\theta_{ci}$. $<r(\theta_{ci}, L_{sc})>$ is calculated using Monte Carlo simulation from a medium with $l_s^* = 2$ mm, and $g = 0.9$ (at $\lambda = 520$ nm) for $\theta_{ci}$ varying from 1° to 90° and different $L_{sc}$ varying between 35 and 140 μm. $<r(\theta_{ci}, L_{sc})>$ is proportional to both $\theta_{ci}$ and $L_{sc}$.

In order to model LEBS, it is necessary to accurately determine the angle at which the photons are collected. The LEBS signals obtained from simulation are collected at $\theta_{ci} \sim 1.5°$ which is close to the width of the experimentally recorded LEBS peaks. This is due to the fact that the angular range of the photons traveling in the time reversed direction restricts the exit angle of the photons in the LEBS experiments. It can be seen from Fig. 6 that the probability of exit angle, $P_{ex}(\theta_{ci})$ converges at small angles of around 1°-3° when spatial coherence light source is used. The LEBS peak width obtained from exit angles, $\theta_{ci}$ ~1.5 degrees deviates less than 5% when compared to the photons collected at low exit angles (0.3° to 1°). We take advantage of this convergence at small angles for better averaging of $P(r)$. However, any change in the exit angle of the photons beyond this would significantly change the depth from which the time reversed photons are obtained leading to erroneous modeling of LEBS. We obtain $P(r)$ at 1.5° using Monte Carlo simulation and then multiply it by the spatial coherence function (Eq. 4) corresponding to the coherence length used in the experiment. The Fourier transform of this multiplied product is then taken to obtain LEBS peak. In subsequent sections, we present detailed validation of the LEBS peak obtained using Monte Carlo simulation by comparing with analytical model and LEBS experiments.

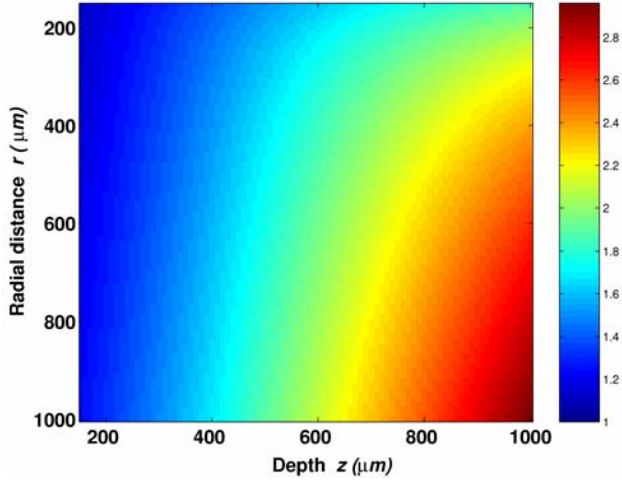

FIG. 5. $<r(\theta_{ci}, L_{sc})>$ as a function of $r$ and depth of penetration of scattered photons $z$ for a fixed $\theta_{ci} = 45°$ and a fixed $L_{sc} = 600$ μm. $<r(\theta_{ci}, L_{sc})>$ is calculated from a medium with $l_s^* = 2$ mm, and $g = 0.9$ (at $\lambda = 520$ nm) using Monte Carlo simulation. $<r(\theta_{ci}, L_{sc})>$ is proportional to the penetration depth $z$.

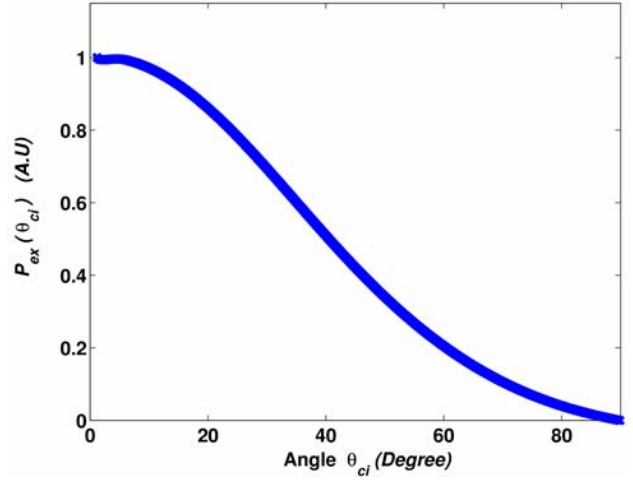

FIG. 6. Probability of exit angle $P_{ex}(\theta_{ci})$ as a function of $\theta_{ci}$ obtained for low order scattering regime. $P_{ex}(\theta_{ci})$ is calculated using Monte Carlo simulation ($l_s^* = 2$ mm, $g = 0.9$ at $\lambda = 520$ nm) for a fixed $L_{sc} = 200$ μm by varying $\theta_{ci}$ between 1° and 90°. $P_{ex}(\theta_{ci})$ converges at small angles of around 1°-3° when $L_{sc} \ll l_s^*$.

### C. Validation of LEBS Monte Carlo Simulations

In order to verify the accuracy of the Monte Carlo simulation, we first compare the profile of the EBS peak obtained from numerical simulation with the profile of the EBS peak calculated by Akkermans *et al.*[40] for an isotropic scattering medium. The simulation is performed with $l_s^* = 2$ mm and infinite spatial coherence ($L_{sc} = 50$ mm), with the exit angle, $\theta_{ci} \approx 1.5°$. Figure 7 shows that the simulation results match well with the analytical results. In order to verify the effect of exit angles in EBS measurements, the simulations from small exit angles, $\theta_{ci} \approx 1.5°$ are compared to those obtained from higher exit angles (e.g. $\theta_{ci} = 80°$). As shown in Fig. 7, the Monte Carlo simulations for smaller exit angle (circles) agrees with the results obtained from large exit angles (asterisks).

We next validate the spatial coherence function (Eq. (4)) used in simulating the LEBS peak by comparing the profiles obtained from the simulation with those of the LEBS experiments (discussed later) for a sample with $l_s^*$ smaller than $L_{sc}$. In this regime ($l_s^* \ll L_{sc}$), though $L_{sc}$ is small, LEBS profiles are primarily determined by $l_s^*$ as the light scattering paths are not affected by the $L_{sc}$ and hence the role of the finite $L_{sc}$ is virtually insignificant.

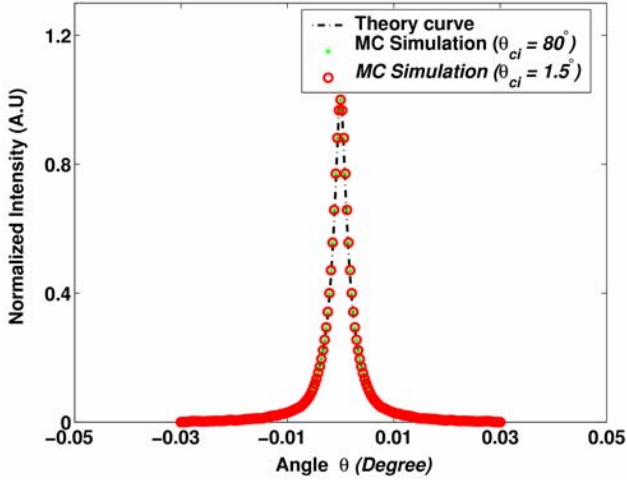

FIG. 7. Comparison of the EBS profile from Monte Carlo simulation with that from analytical formulation.[40] The profile of EBS peak from Monte Carlo simulation $I_{EBS}(\theta)$, is calculated from a medium with $l_s^* = 2$ mm, and $g = 0.9$ (at $\lambda = 520$ nm) for $L_{sc} = 50$ mm. The results from simulation are in excellent agreement with the analytical results. Also, the LEBS simulation for $\theta_{ci} = 1.5°$ agrees well with the results obtained from $\theta_{ci} = 80°$ as the width of the EBS peak is insensitive to the $\theta_{ci}$ in the diffusive multiple scattering regime ($L_{sc} \gg l_s^*$).

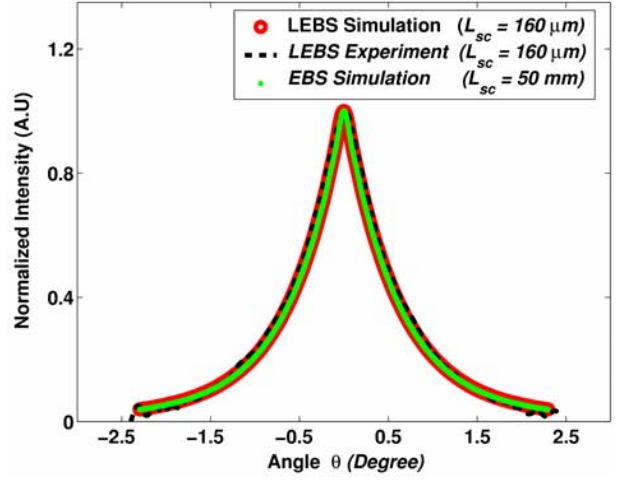

FIG. 8. Normalized intensity profile of the LEBS peak as a function of $\theta$ from Monte Carlo simulation is compared with that of the experimental result from white paint under low spatial coherence illumination (Xenon lamp, $\lambda = 520$ nm, $L_{sc} = 160$ μm). $I_{LEBS}(\theta)$, is calculated using Monte Carlo simulation from a medium with $l_s^* = 4$ mm, and $g = 0.9$ (at $\lambda = 520$ nm) for $L_{sc} = 160$ μm. The simulation results agrees well with experimentally observed LEBS peak when $L_{sc} = 160$ μm. Also, the EBS peak from $L_{sc} = 50$ mm and LEBS peak from $L_{sc} = 160$ μm are completely indistinguishable as the peak width is completely determined by the small $l_s^*$ of the medium and the spatial coherence length plays an insignificant role in this regime.

We simulate the LEBS peak with $L_{sc} = 160$ μm from a sample of $l_s^* = 4$ μm. The LEBS peak in the experiment is collected from white paint (Benjamin Moore) with $l_s^* = 4$ μm under low spatial coherence illumination (Xenon lamp). As shown in Fig. 8 the LEBS peak obtained from Monte Carlo simulation matches well with the experimentally observed LEBS peak for $l_s^* \ll L_{sc}$. This validates the use of spatial coherence function in the modeling of LEBS peak for low spatial coherence illumination. We further compare the LEBS profile with the EBS peak obtained from infinite spatial coherence length ($L_{sc} = 50$ mm). As expected, the EBS and LEBS peaks are completely indistinguishable as the peak width is completely determined by the small $l_s^*$ of the medium (Fig. 8). In the next section, we present the experimental verification of the LEBS simulation in the low order scattering regime ($L_{sc} \ll l_s^*$).

### D. Comparison of Monte Carlo simulation with the experimental results

The LEBS peak obtained using Monte Carlo simulations are experimentally verified. The detailed description of our experimental setup is given elsewhere.[23,24] Here we describe essential parts in brief. The experimental setup consists of a 500-W Xe lamp (Oriel) to deliver a continuous wave broadband light which is then collimated using a 4-f lens system. The collimated light is polarized and delivered onto a sample at an incident angle of 15° to prevent specular reflection. The spatial coherence length of illumination $L_{sc}$ is varied between 30 and 220μm, which was confirmed by a double-slit interference experiment.[34] The backscattered light from the sample is sent through a setup consisting of a Fourier lens and a polarizer oriented along the same direction of the incident light. The co-polarized light is then collected by an imaging spectrograph (Acton Research) positioned in the focal plane of the Fourier lens and coupled to a CCD camera (CoolsnapHQ, Roper Scientific). The angular distribution of the backscattered light from the sample is projected onto the slit of the spectrograph which disperses the light according to the wavelength in the direction perpendicular to the slit. The CCD camera records a matrix of light scattering intensity as a function of backscattering angle $\theta$ at different wavelengths $\lambda$. In each CCD pixel, the collected light is integrated within a narrow band of wavelengths around $\lambda$, with the width of the band determined by the width of the spectrograph slit. The LEBS peaks are normalized by the incoherent baseline measured at large backscattering angles ($\theta > 4°$). The resulting LEBS signal is compared with the signals predicted by Monte Carlo simulation.

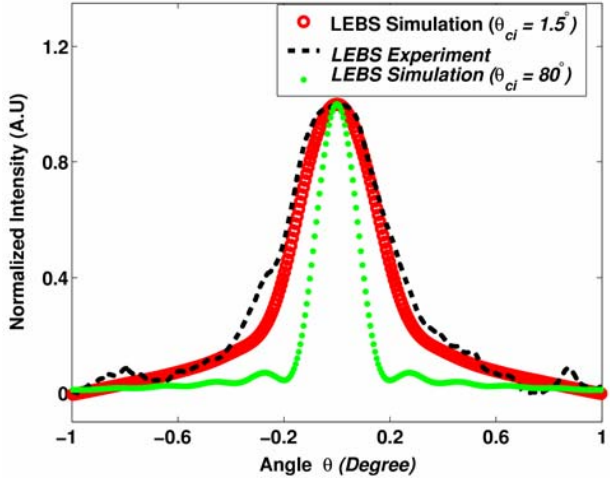
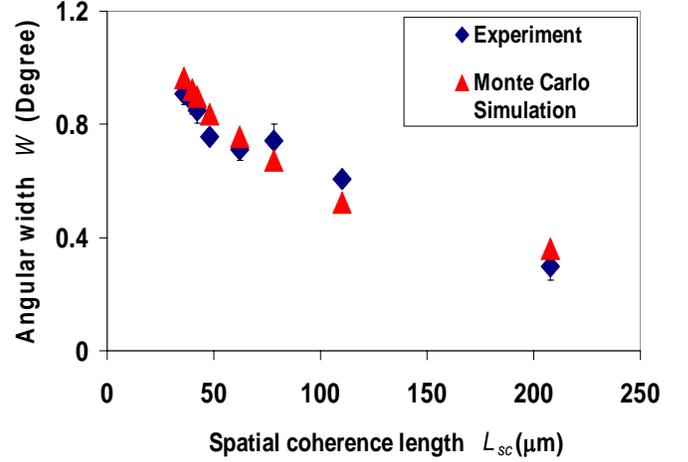

FIG. 9. Normalized intensity profile of the LEBS peak as a function of angle $\theta$ from Monte Carlo simulation is compared with that of the aqueous suspensions of polystyrene microspheres (diameter = 0.89 μm) under low spatial coherence illumination (Xenon lamp, $\lambda$ = 520 nm, $L_{sc}$ = 42 μm). $I_{LEBS}(\theta)$, is simulated for a medium with $l_s^*$ = 2 mm, $g$ = 0.9 (at $\lambda$ = 520 nm) and $L_{sc}$ = 42 μm. LEBS peak simulated by Monte Carlo simulation at $\theta_{ci}$ = 1.5° matches well with the LEBS peak profile recorded in the experiment. On the contrary, the LEBS peaks from $\theta_{ci}$ = 80° are three times narrower than those obtained from the experiment and it does not accurately predict the LEBS peak as it is insensitive to low orders of scattering.

FIG. 10. Comparison of angular width $W$ of the LEBS peaks obtained from Monte Carlo simulation with that of LEBS peaks from experiment (bead diameter = 0.89μm) under low spatial coherence illumination (Xenon lamp, $\lambda$ = 520 nm). The width of LEBS peak from simulation ($l_s^*$ = 2 mm, $g$ = 0.9 at $\lambda$ = 520 nm) is calculated for 8 different $L_{sc}$ varying between 30 μm and 220 μm at a fixed $\theta_{ci}$ = 1.5°. The error bars in the curves are the standard errors. The widths of the LEBS peaks predicted by Monte Carlo simulation are in excellent agreement with those determined from the experiment.

We record LEBS from aqueous suspensions of polystyrene microspheres (Duke Scientific, Palo Alto, CA) of various diameters from 200 nm to 890 nm. The dimension of the samples is $\pi \times 50^2$ mm² $\times$ 100 mm. We vary the transport mean free path $l_s^*$ from 500 μm and 2000 μm and the spatial coherence lengths from 30 μm and 220 μm. As a representative, we show here the results obtained from the sample of $l_s^*$ = 2000 μm and bead diameter of 0.89 μm. As shown in Fig. 9, LEBS peak predicted by Monte Carlo simulations is in excellent agreement with the peak recorded in the experiment ($L_{sc}$=42 μm). Further, the LEBS peak obtained from higher exit angles (asterisk) does not accurately model the LEBS peak as it is insensitive to the low orders of scattering. On the other hand, the broadening of LEBS peak can be accurately obtained only at low exit angles (circle) as it is more sensitive to the low orders of scattering. The simulations are further verified for eight other values of $L_{sc}$ varying from 35 to 220 μm (Fig. 10). The error bars in the curves are the standard error obtained from 3 different sets of experiments conducted at different spatial coherence lengths. As shown in Fig. 10, the widths of LEBS peaks predicted by the Monte Carlo simulations are in excellent agreement with those determined in the experiment. This confirms that the Monte Carlo simulation can be used to model LEBS.

**4. Conclusion**

In conclusion, we have demonstrated for the first time that the photon random walk model can be used to model low coherence enhanced backscattering from a weakly scattering medium. We have shown that LEBS predicted by the simulations are in excellent agreement with the experimental data. Furthermore, we have demonstrated that the exit angles of the photons, which are typically neglected in modeling of conventional EBS, play key role in modeling of LEBS. As EBS peaks depend on the Fourier transform of the radial distribution of the intensity on the surface of the sample, it is extremely important to consider the change due to exit angle of the photons in the shape of probability distribution *P(r)*. Our data indicate that *P(r)* obtained from low order scattering is sensitive to the exit angles of the photons, which in turn depend on the depth from which the photons are collected

**Acknowledgement**

This study was supported in part by National Institutes of Health grant, R01 EB003682, R01 CA112315 and National Science Foundation grant BES-0238903.